# Sunspot Observations by Hisako Koyama: 1945 – 1996

Hisashi Hayakawa (1-2)*, Frédéric Clette (3)**, Toshihiro Horaguchi (4), Tomoya Iju (5), Delores J. Knipp (6-7), Huixin Liu (8), Takashi Nakajima (4)

(1) Graduate School of Letters, Osaka University, Toyonaka, 5600043, Japan (JSPS Research Fellow).
(2) UK Solar System Data Centre, Space Physics and Operations Division, RAL Space, Science and Technology Facilities Council, Rutherford Appleton Laboratory, Harwell Oxford, Didcot, Oxfordshire, OX11 0QX, UK
(3) World Data Center SILSO, Royal Observatory of Belgium, 3 avenue Circulaire, 1180 Brussels, Belgium
(4) National Museum of Nature and Science, Tsukuba, 3050005, Japan
(5) National Astronomical Observatory of Japan, Mitaka, 1818588, Japan.
(6) High Altitude Observatory, National Center for Atmospheric Research, Boulder, CO, CO 80307, USA
(7) Smead Aerospace Engineering Sciences Department, University of Colorado Boulder, Boulder, CO, CO 80309, USA
(8) FFaculty of Science, Department of Earth and Planetary Science, Kyushu University, Fukuoka, 8190395, Japan

* hayakawa@kwasan.kyoto-u.ac.jp; hisashi.hayakawa@stfc.ac.uk
** frederic.clette@oma.be

**Abstract**

The sunspot record is the only observational tracer of solar activity that provides a fundamental, multi-century reference. Its homogeneity has been largely maintained with a succession of long-duration visual observers. In this paper, we examine observations of one of the primary reference sunspot observers, Hisako Koyama. By consulting original archives of the National Museum of Nature and Science of Japan (hereafter, NMNS), we retrace the main steps of her solar-observing career, from 1945 to 1996. We also present the reconstruction of a full digital database of her sunspot observations at the NMNS, with her original drawings and logbooks. Here, we extend the availability of her observational data from 1947-1984 to 1945-1996. Comparisons with the international sunspot number (version 2) and with the group sunspot number series show a good global stability of Koyama's observations, with only temporary fluctuations over the main interval 1947-1982. Identifying drawings made by alternate observers throughout the series, we find that a single downward baseline shift in the record coincides with the partial contribution of replacement observers mostly after 1983. We determine the correction factor to bring the second part (1983-1996) to the same scale with Koyama's main interval (1947-1982). We find a downward jump by 9% after 1983, which then remains stable until 1996. Overall, the high quality of Koyama's observations with her life-long dedication leaves a lasting legacy of this exceptional personal achievement. With this comprehensive recovery, we now make the totality of this legacy directly accessible and exploitable for future research.

**1. Introduction**

The sunspot number is arguably the most standard index to evaluate the variable solar activity, and it is one of the most studied time series in astrophysics (Charbonneau, 2010; Clette *et al*., 2014; Hathaway, 2015). The sunspot record starts in 1610 and forms one of the longest ongoing scientific experiments (Vaquero, 2007; Vaquero and Vazquez, 2009; Owens, 2013; Clette *et al*., 2014; Vaquero *et al*., 2016; Usoskin, 2017; Muñoz-Jaramillo and Vaquero, 2019). This global solar activity index is used in a wide range of scientific research and technological applications, such as the solar dynamo (Charbonneau, 2010; Auguston *et al*., 2015; Hotta *et al*., 2016, 2019), solar grand minima (Vaquero *et al*., 2011, 2015; Usoskin *et al*., 2015), space climate (Hathaway and Wilson, 2004), space weather (Pulkkinen, 2007; Hayakawa *et al*., 2018b, 2019a), solar-terrestrial relations (Lockwood, 2013; Hayakawa *et al*., 2018a, 2019b), influence to terrestrial climate (Lockwood, 2012; Owens *et al*., 2017), and solar cycle prediction (McNish and Lincoln, 1949; Svalgaard *et al*., 2005; Cameron *et al*., 2016; Hathaway and Upton, 2016; IIjima *et al*., 2017).

Presently, two time series are derived from visual sunspot counts: the international sunspot number (ISN) and the group sunspot number (GSN) (Clette *et al*., 2014, 2015). The ISN, initiated in 1849 by R. Wolf, is defined as $10 N_g + N_s$, where $N_g$ is the total number of sunspot groups and $N_s$ is the total number of spots visible on the solar disk. The ISN thus takes into account both the number of sunspot groups and the number of individual spots, which gives a measure of the size of the groups and thus of the strength of the underlying magnetic fields (Clette *et al*., 2014; Clette and Lefèvre, 2016). The GSN has been introduced more recently by Hoyt and Schatten (1998a, 1998b) and is only based on group counts $N_g$, with all groups counted as 1 regardless of their very diverse sizes. The GSN is thus a cruder sunspot index but it has the advantage to be derivable even from very early observations, and hence, it allows reconstructing solar activity before the mid-18$^{th}$ century and back to the first telescopic observations in 1610 (Clette *et al*., 2014).

While these two similar indices agree with each other after 1900, they show considerable discrepancies in their amplitude, before ~ 1900. This disagreement leads to incompatible interpretations of the evolution of solar activity before 1885 (Cliver, 2016, 2017; Cliver and Ling, 2016; Usoskin, 2017). Therefore, a thorough revision of the calibration of both series is currently ongoing on the basis of original observations and using new modern methodologies (Clette *et al*., 2014, 2015; Clette and Lefèvre, 2016; Vaquero *et al*., 2016; Svalgaard and Schatten, 2016; Usoskin *et al*., 2016; Cliver and Ling, 2016; Chatzistergos *et al*., 2017; Muñoz-Jaramillo and Vaquero, 2019). While observations were conducted over many decades in various institutional observatories such as Greenwich (Willis *et al*., 2013, 2016a, 2016b), Debrecen (Baranyi *et al*., 2016; Győri *et al*., 2017), or Mount Wilson (Lefèvre *et al*., 2005; Pevtsov *et al*., 2019), long-term observations at smaller observatories by individual astronomers – frequently skilled amateur observers – are of particular importance in terms of stability. Indeed, such sunspot time series have been derived from a single observer, often with a unique telescope, in contrast with professional teams working in shifts and with a staff changing over the years (*e.g*., Clette *et al*., 2014; Carrasco *et al*., 2019).

Among them, the detailed sunspot observations by Hisako Koyama (1916 – 1997) of the National Science Museum (NSM; current NMNS) are known as one of the longest and most stable reference data series available during the 20$^{th}$ century (Horaguchi and Nakajima, 2001; Clette *et al*., 2014; Knipp *et al*., 2017). Koyama was chosen as one of the reference observers in the recent so-called "backbone" reconstruction of the GSN (Clette *et al*., 2014; Svalgaard and Schatten, 2016). Her work stands with other prominent historical solar observers: Staudach (Arlt *et al*., 2008; Svalgaard, 2017), Schwabe (Arlt *et al*., 2013), Wolfer (Wolfer, 1907), and Cortesi (Cortesi *et al*., 2016; Clette *et al*., 2016). At the occasion of this GSN revision, Svalgaard and Schatten (2016) examined her group counts as published in Koyama's main publication (Koyama 1985) and pointed out a decrease of her counting scale after 1981, interpreting it as a loss of visual acuity. For their GSN reconstruction, Svalgaard and Schatten (2016) independently digitized the group numbers listed in Koyama (1985). These are the numbers that were included in the latest group number database (Vaquero *et al*., 2016).

Ms. Hisako Koyama published almost all her observations from 1947 to 1984 in a monograph (Koyama, 1985). Moreover, her original drawings have been preserved in the National Museum of Nature and Science of Japan, and their digitization has been started by Horaguchi and Nakajima (2001). Based on their survey, Horaguchi and Nakajima (2001) could confirm that Hisako Koyama continued her sunspot observations well after the publication of her monograph (Koyama, 1985), for the rest of her life until early 1997. Her life and scientific contributions were recently brought to wider attention by Knipp *et al*. (2017), raising further scientific interest.

In this article, we first provide an overview of Hisako Koyama's sunspot observations, and we chronicle of her observing activities. We then highlight the recent release of the digital database by the NMNS[1] since 2017, following the pilot efforts by Horaguchi and Nakajima (2001). Finally, we compare Hisako Koyama's personal sunspot numbers with the multi-station reference series of the international sunspot number maintained by the World Data Center "Sunspot Index and Long-term Solar Observations" (SILSO) to check for any scale change over the years and their possible coincidence with known events in her long observing career.

**2. Hisako Koyama's Observational Records:**

Koyama's observations made between 1947 and 1984 were published in her monograph (Koyama, 1985) and have been converted to digital form after the pilot efforts by Horaguchi and Nakajima (2001). Her original sunspot drawings are preserved in the NMNS at Tsukuba. Her sunspot observations were also summarized for every month in her own handwritten logbooks. Koyama also sent monthly reports to scientific institutes such as the Swiss Federal Observatory in Zürich and the Royal Observatory of Belgium in Brussels (Koyama, 1981, p.114; O'Meara, 1987; Fujimori, 1994; Knipp *et al.*, 2017).

---

[1] https://www.kahaku.go.jp/research/db/science_engineering/sunspot/

While Koyama's monograph (Koyama, 1985) gathers the results in volumes 1 to 38 of her handwritten logbooks (hereafter, we abbreviate "volume" as "v."), she compiled more logbooks (v.39 – v.50) until 1996 Dec. 31. Koyama thus continued her observations over a full solar cycle after the publication of her 1985 compilation. This final part of her observations remained unpublished prior to this digitization (see Figure 1).

In the archive at Tsukuba, her sunspot drawings are wrapped by year, and categorized according to the volumes of her logbooks. This drawing archive also includes earlier observations made before January 1947, when her first logbook was compiled. At the NMNS, we have now also located her earliest series of sunspot drawings: the very first drawings spanning the period from 1945 Sep. 6 to 1946 Jan. 23, made with her personal 3-cm aperture telescope before coming to Ueno, and a first series of drawings made at Ueno between 1946 May 4 and 1947 Jan. 21 with a 20-cm refracting telescope and projection screen, then still with a rather small solar disk diameter of only 10-cm (see Figure 2). Years later, in November 1975, Koyama compiled a notebook for the year 1946, with this first series of observations made at Ueno, as volume 0. However, this early part of her observations was not included in her 1985 compilation, because it was obtained with different telescopes, as explicitly mentioned in her logbook v. 1.

Hayakawa et al. (2020) Sunspot Observations by Hisako Koyama: 1945 – 1996, *Monthly Notices of the Royal Astronomical Society*. DOI: 10.1093/mnras/stz3345

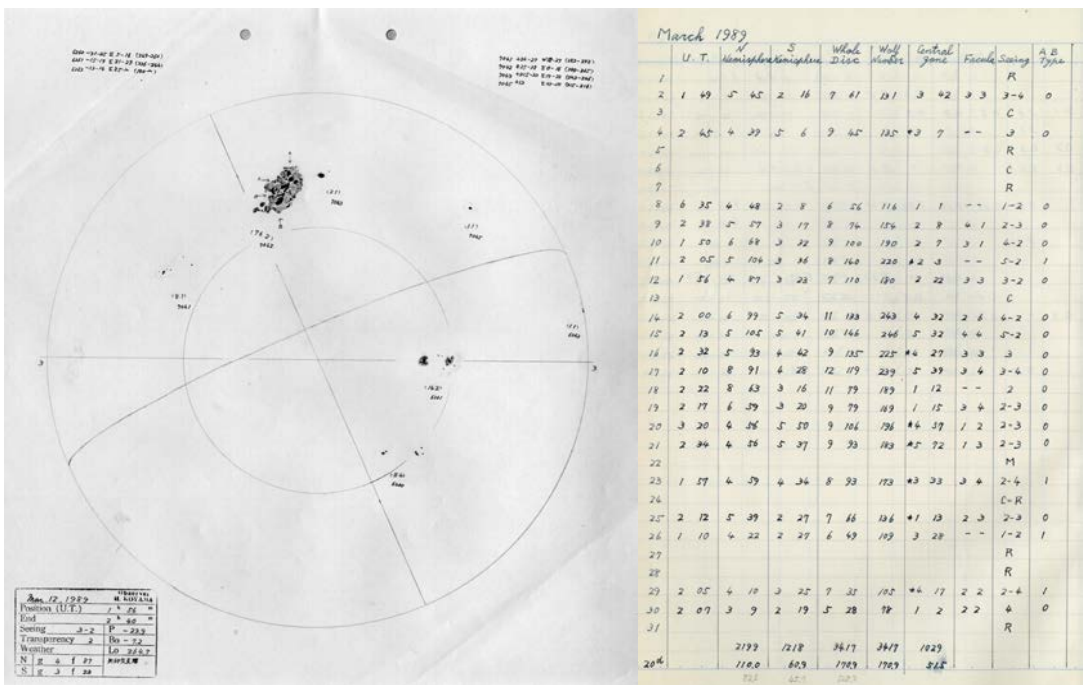

Figure 1: Examples of Koyama's late sunspot drawings and logbooks; (left) sunspot drawing for 1989, March 12, and (right) logbook page for March 1989 (courtesy of the National Museum of Nature and Science of Japan)

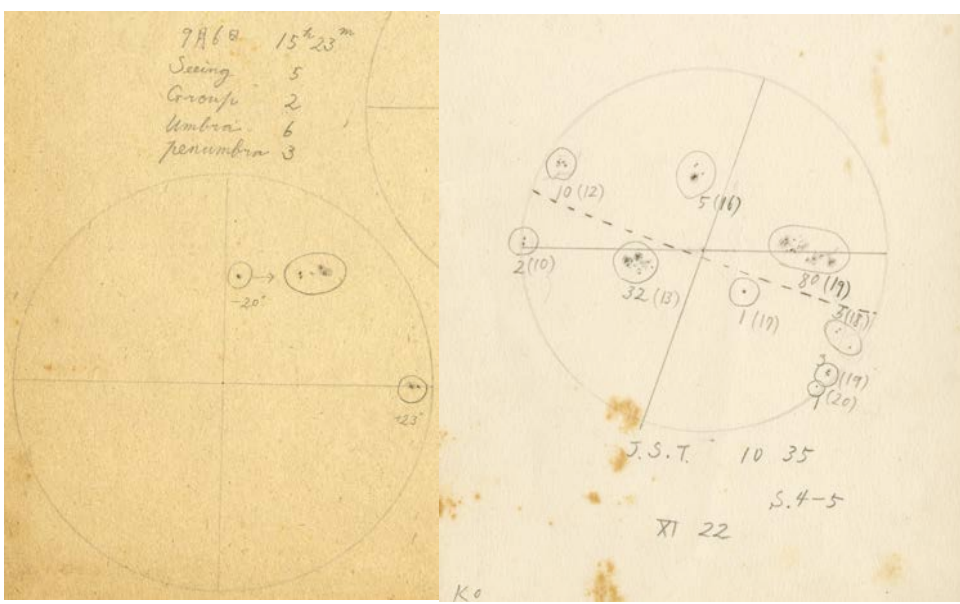

Figure 2: Examples of Koyama's early sunspot drawings: (left) Koyama's sunspot drawings for 1945 Sep. 6, made with the 3-cm aperture telescope; (right) Koyama's sunspot drawing for 1946 Nov. 22, made with the 20-cm refractor and 10-cm projection.

## 3. Koyama's Telescopes

During her 51-year long observing career, Koyama used several different telescopes, as summarized in Table 1. Her early observations before 1947 were not included in her first logbooks nor in the 1985 monograph (Koyama, 1985). As described in her logbook v. 1., Koyama indicated her main telescope for each year on the wrapping paper containing her sunspot drawings, and on the front covers of her yearly logbooks. Moreover, on her daily sunspot drawings, she mentioned when she used alternate telescopes or asked a colleague to observe in her absence. According to her notes on the wrapping paper and front covers, she first used eyepiece observations (aperture = 3 cm) from 1945 Sep. 6 to 1946 Jan. 23, and switched to the projection method with a larger refractor (aperture = 20 cm) and a 10-cm solar disk from 1946 May 4 to 1947 Jan. 21.

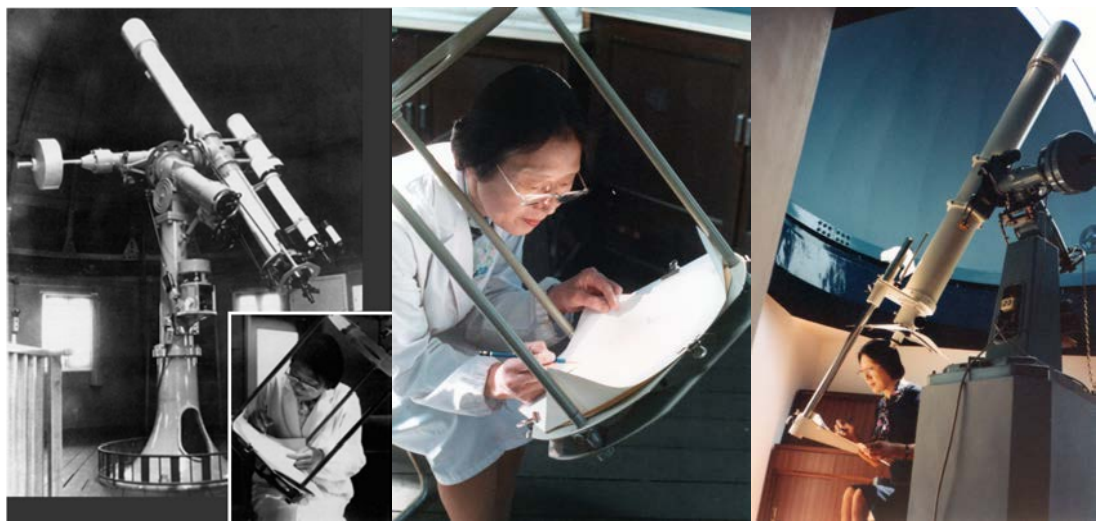

Figure 3: (left) Hisako Koyama and the 20-cm refracting telescope with the projection screen at the NSM at Ueno Park (N35°43′, E139°46′), reproduced from Koyama (1985) with an enlarged-image inset; (centre) Hisako Koyama using the 30-cm projection screen at NSM (courtesy of Noji Collection); (right) Hisako Koyama and the 15-cm refractor that she used in Murayama's household (N35°43′, E139°45′) (courtesy of Noji Collection).

As the 10-cm projected solar disk was quite bright with the 20-cm refractor, Koyama adapted the projection screen for a larger 30 cm image diameter, following the recommendations of Masaaki Furuhata (1912 – 1988) of Tokyo Astronomical Observatory. From then onward, thus from 1947 Jan. 22 to 1988, she almost exclusively used this new instrument configuration, which can thus be considered as the standard reference instrument for Koyama's series of sunspot observations. (Figure 3; see Koyama, 1985). This telescope is currently exhibited at the NMNS at Ueno.

During Koyama's epoch, the research division of the NSM was located at Ueno, where she conducted her daily observations in a rooftop dome (Figures 4 and 5; Koyama, 1985; Knipp *et al.*, 2017). The observatory dome (Figures 4 and 5) was erected in 1931, as part of the construction of the museum, which was called the NSM at that time (National Science Museum, 1977, p. 315). As the observatory was within walking distance from Koyama's home, she could easily go to the observatory at any time, even outside her working days (Koyama, 1981). This favorable location also helped her to continue her observations for a considerable period, even at an advanced age. Later on, the research division moved to Tsukuba in 2011 via Okubo, and the telescope was moved to museum exhibition at Ueno.

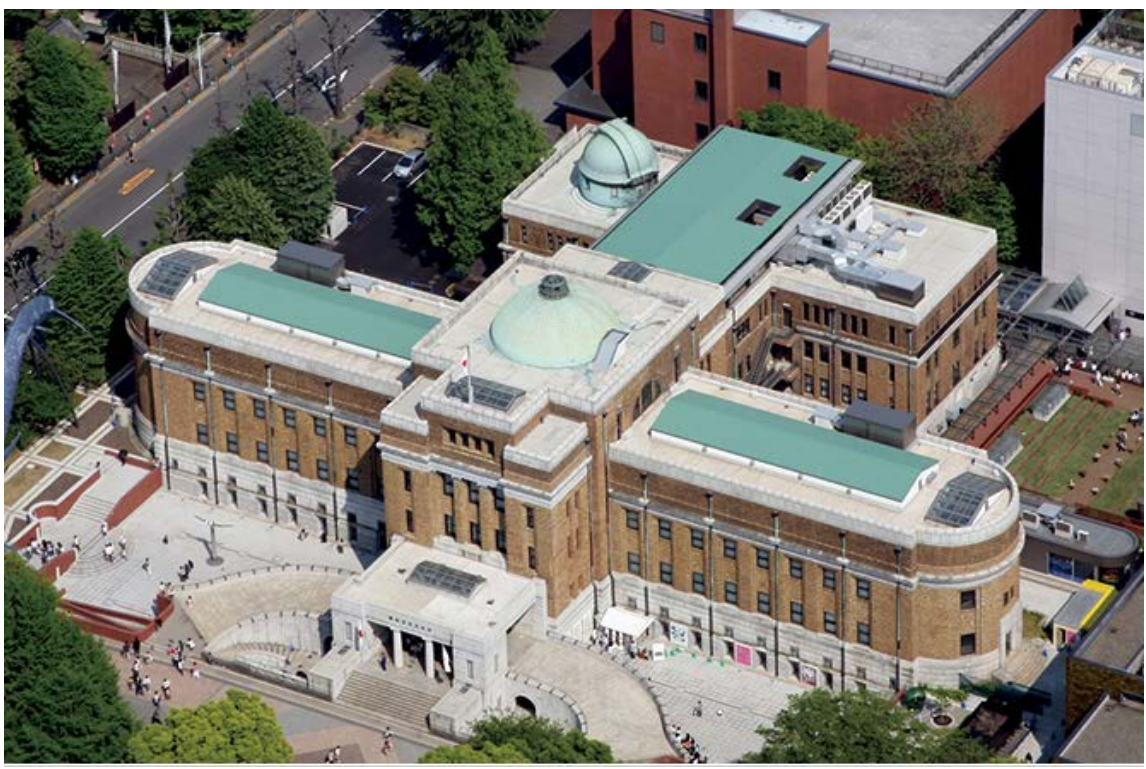

Figure 4: The main building of the NMNS with the roof-top copper-coloured

observatory dome in the NE of this building (top center). The museum building was produced with a design of airplane heading westward (Tokyo Science Museum, 1931). This view is from the West towards the East. The Museum is surrounded by other large buildings on its East (the Japan Academy) and South (the new museum building), and by the large Ueno Park on the West. (Image courtesy of the National Museum of Nature and Science).

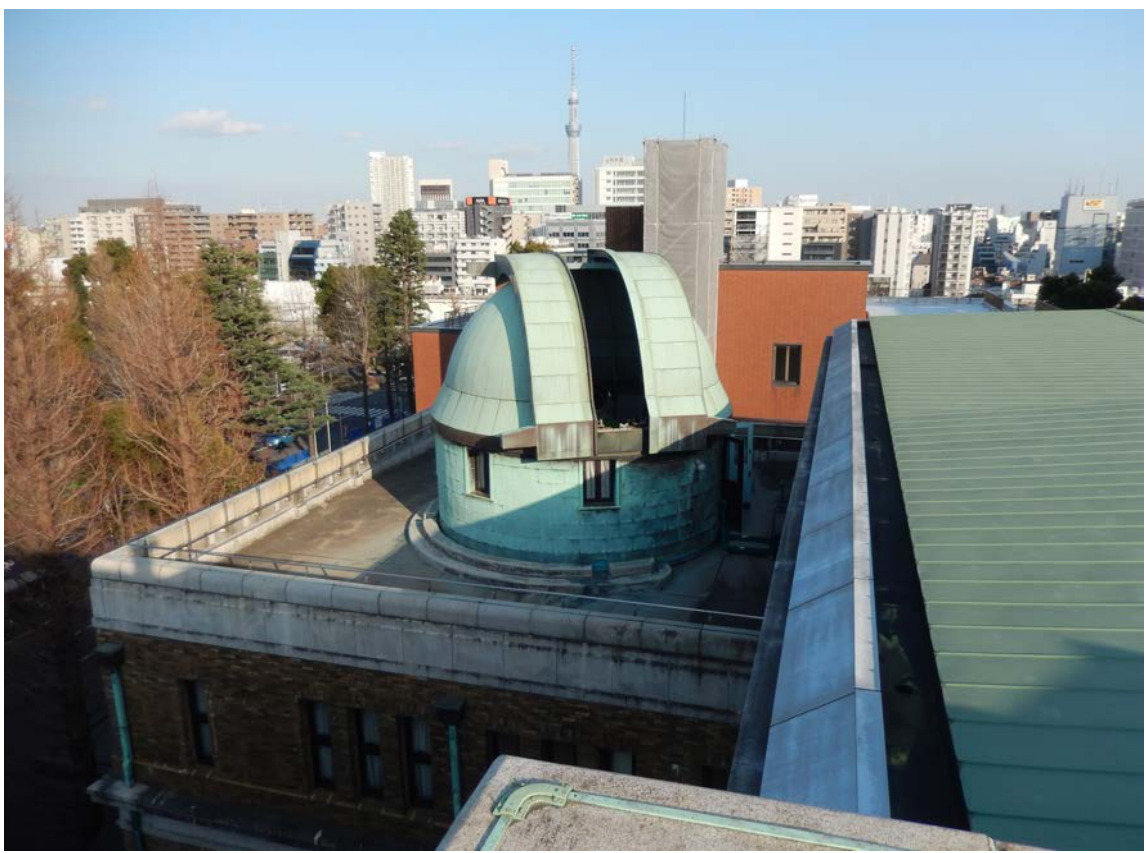

Figure 5: Close-up view of the observatory dome photographed in March 2018. Here, the dome slit is oriented towards the West. The copper-coloured roof on the South of the observatory dome was newly constructed in 2006, and hence did not exist at the time of Koyama's observations. The surrounding wall remained unchanged since Koyama's epoch and still wears graffiti made by school children who visited this dome between c. 1950 and 1970.

She continued using this telescope as a museum fellow, after her official retirement in March 1981 (Koyama, 1985). From 1989 to 1991, she transitioned from this 20-cm refractor with 30 cm projection to a 15-cm refractor with 30 cm projection in Murayama's household. Sadao Murayama (1924 – 2013) was director of the Department of Science and Engineering until March 1989 and collaborated with Koyama throughout her observational career. Finally, Koyama settled for the 15-cm refractor for her last observations from April 1991 to 1996 (see also Fujimori, 1994, p.45).

Apart from these main telescopes, Koyama occasionally used a few other telescopes such as a portable 8-cm refractor, when the main telescope was not available, *e.g.* during repairs or maintenance of the observatory dome, as recorded in her logbooks. As these observations are different in quality from her main-telescope observations, we did not use those observations in our analyses.

Table 1: Chronology of the main telescopes used by Koyama.

| Chronology | Telescope | Method |
| --- | --- | --- |
| 1945 Sep. – 1946 Jan. | 3-cm refractor | eyepiece observation |
| 1946 May – 1947 Jan. | 20-cm refractor | 10-cm projection |
| 1947 Jan. 22 – 1991 | 20-cm refractor | 30-cm projection |
| 1989 – 1996 | 15-cm refractor | 30-cm projection |

**4. Observational Results:**

Based on her solar observations, in her logbooks, Koyama registered numerical tables of sunspot numbers, and also synoptic tables retracing the day-to-day evolution of sunspot groups. In the tables of sunspot numbers, she provided the date, the observing time in Universal Time (UT), the group count (g) and spot count (f) for the northern hemisphere, the southern hemisphere, the whole visible disk, and the central circular zone (half solar diameter), the Wolf number for the whole visible disk and the central circular zone, the seeing quality, and notes. In the synoptic tables recording the evolution of sunspot groups, she provided the group number, the heliographic latitude, and longitude, the date of central meridian passage, the dates of first observation and last observation, the evolution of the group in the Zürich group-type classification, and notes. The digital database follows exactly the same format as the original paper documents.

Figure 6 shows the number of Koyama's observing days for each year. Koyama consistently observed the solar disk for more than 150 days every year from 1947 to 1996. We can see the progressive transition between her two main telescopes, from 1989 to 1991. Even in her last years, she continued observing intensively for more than 150 days/year, and this observing rate shows hardly any dependence on the solar cycle evolution. As stated by Koyama herself, she was diligent enough to visit the observatory whenever the weather permitted: "It takes 30 minutes from my home to the museum, by a fast walk. I have been happy enough to be free from either packed trains or traffic jams. I have come for observing even on holidays, whenever it was sunny" (Koyama, 1981, p.113)

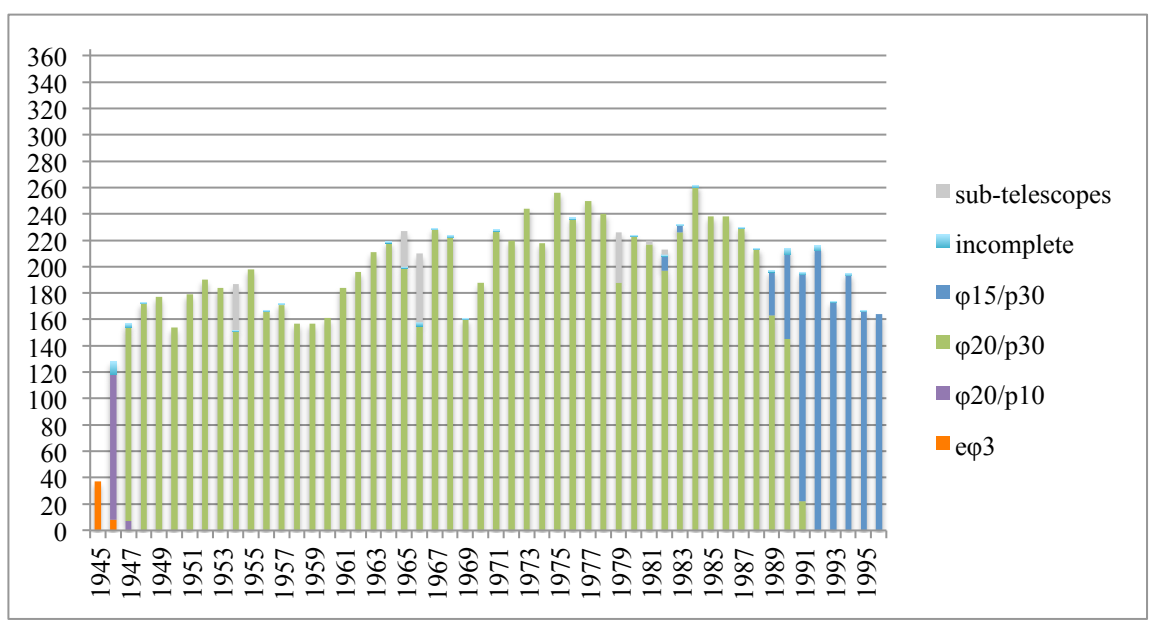

Figure 6: Number of Koyama's observing days per year. The aperture size is indicated with φ. The observations with the 3-cm telescope and by eyepiece observations are shown in orange, those by the 20-cm refractor are shown in green (30-cm projection) and purple (10-cm projection). Those made with the 15-cm refractor are shown in dark blue, and those by other telescopes are shown in gray.

Nevertheless, Koyama concedes that her observations were not completely immune from weather conditions. According to her, the worst condition was "not rain, but a day with clouds frequently coming and going". On such days, she often had to stay for hours in the observatory, waiting for a possible break in the clouds. With humor, she once reflected that she could "probably win a competition for endurance, if there was one" (Koyama, 1981, p.113; Knipp *et al*., 2017). Occasionally, she was even unable to complete an observation due to the cloud cover, which left some reports incomplete (Koyama, 1981, p.113). In the archives, some observations with missing details can clearly be categorized among such interrupted observations.

With this comprehensive and dedicated series of observations, Koyama managed to accumulate a unique long-term sunspot record spanning roughly half a century. Figure 7 shows the corresponding monthly mean total Wolf number over the whole visible disk, as well as the monthly-mean hemispheric Wolf number for the northern and southern hemispheres.

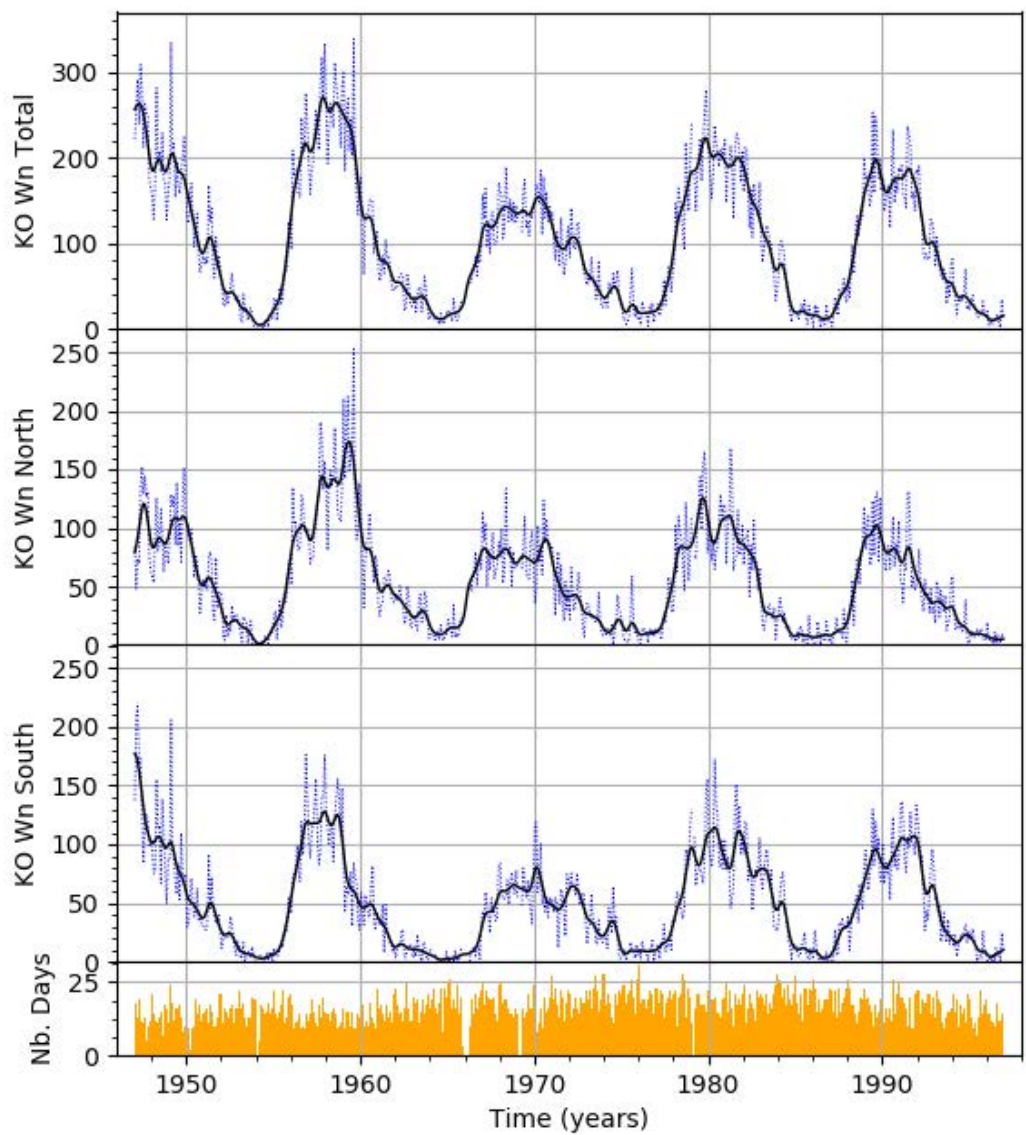

Figure 7: Plot of the monthly-mean Wolf numbers (fine dotted lines), for the whole Sun (Total, top panel) and for each hemisphere (North, second panel; South, third panel). In each panel, the solid curve was obtained by a Gaussian smoothing of the monthly means with a width at half-maximum of 7 months, in order to better show the long-term evolution. The number of observed days per month is plotted in the lower panel. It shows a few gaps lasting for at least one full month.

Her hemispheric numbers are of particular interest, as hemispheric sunspot numbers were produced at the World Data Center SILSO only since 1992, on the base of its worldwide network. Monthly mean Wolf numbers from only the Uccle Station[2] in Brussels, Belgium, are currently used to extend this series backwards over the interval 1950 to 1994. As Koyama's data cover a similar period, they can thus bring additional data to extend hemispheric numbers back in time before 1992, adding 4 solar cycles. Such hemispheric information will benefit the validation of the current theoretical models of the solar dynamo, which now involve two loosely-tied dynamos, each developing in one solar hemisphere (Charbonneau, 2010).

**5. Homogeneity of Koyama's observations**

**5.1 Comparison with the latest international sunspot number ($S_N$ Version 2.0)**

By its long-duration and excellent continuity, Koyama's long-term sunspot record forms one of the best reference series for the second part of the 20$^{th}$ century. In particular, Koyama was chosen as one of the so-called "backbone" observers for one of the most recent reconstructions of the sunspot group number (see Clette *et al.*, 2014; Svalgaard and Schatten, 2016). Therefore, it is of prime interest to verify the homogeneity of Koyama's sunspot numbers over the full duration of the series. This means checking if the scale of her personal sunspot and group counts changed significantly over time, paying particular attention to slow trends or step-like jumps separating periods with different stable scales.

Svalgaard and Schatten (2016) found that the standard deviation of Koyama's group counts was about 8%, globally over the entire series. They also pointed out a significant decrease of her counts after ~ 1981, relative to other parallel observers, interpreting this late deviation as a loss of visual acuity. As this analysis only considered group counts, it did not indicate if the counts of individual sunspots were also affected by this apparent drift.

Since its release in 2015 July, the new re-calibrated international sunspot number (Version 2.0) compiled in the World Data Center SILSO offers a new state-of-the-art reference (Clette and Lefèvre, 2016). This reference index is based on a worldwide network of stations, which allowed to eliminate any inhomogeneity or outlying number present only in the data of any of the individual stations, leading to a robust reference with a much higher temporal stability than any single raw series. Koyama's Wolf numbers entered the production of this sunspot number series, but at a much reduced level compared to the group number. During the Zürich era, before 1980, data from external stations were only used occasionally to fill gaps when the Zürich station could not observe the Sun due to bad weather, and Koyama was one of those about 40 auxiliary stations. Quite differently, for the period after 1980, SN version 2 was re-constructed by global statistics over 42 long-duration and high-quality stations. In this case, all stations were thus used for each daily value, but Koyama is then only one out of 42 stations. Therefore, contrary to the recent "backbone" group number by Svalgaard and Schatten (2016), Koyama was not used as a single primary reference. Therefore, SN version 2 is largely independent from the detailed characteristics of Koyama's Wolf number series. It can thus be used as an independent reference.

---

[2] http://www.sidc.be/silso/monthlyhemisphericplot

Figure 8 shows a full comparison of the raw Wolf numbers from Koyama $W_K$ and $S_N$ Version 2. For our calculations, we used monthly mean values, and for the plots we smoothed the resulting curves by a Gaussian function with a width of 7 months at half-maximum, in order to filter out the large random variations at time scales below one year, partly caused by the randomness of solar activity itself over timescales of days and weeks. As a verification, we did the same calculations and regressions based directly on daily values and also on yearly means. All these analyses deliver similar results and lead to identical conclusions.

The top panel shows a very good overall match between $W_K$ and $S_N$. Most of the time, the $W_K$ values are slightly below the $S_N$ values, which is most noticeable for the cycle maxima. This indicates that Koyama was counting slightly less sunspots than the reference observers of the Zürich observatory, who define the unit scale of the international sunspot number. Still, there are a few noticeable mismatches in the early data before 1948, in the middle of the strong Solar Cycle 19, around 1960, and in the last observed Solar Cycle 22.

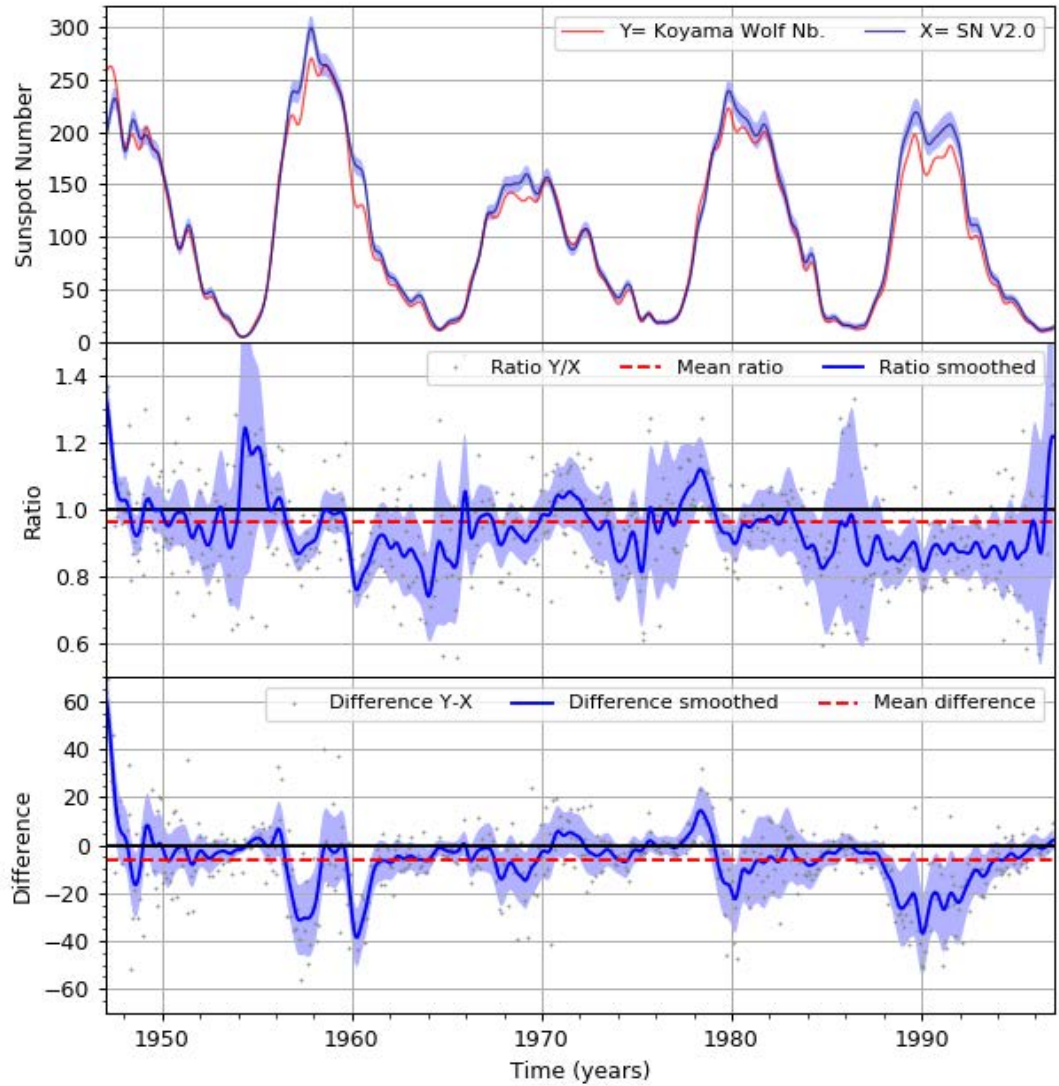

Figure 8: Comparison between the raw Wolf numbers from Koyama $W_K$ and the reference sunspot number series $S_N$ Version 2.0. All curves are smoothed with a Gaussian kernel with a 7-month width at half maximum. The top panel shows the two series: $W_K$ in red and $S_N$ in blue with shading showing the uncertainties on the monthly mean $S_N$ values. The middle and lower panels show the ratio $W_K/S_N$ and difference $W_K$-$S_N$ respectively, with the standard error given by the shaded interval. The individual points are the unsmoothed monthly mean values, and the dashed red lines are the overall mean ratios and differences over the entire 1947 – 1995 time interval.

This is confirmed in the two lower panels, when checking the intervals when the ratio $W_K/S_N$ or difference $W_K$-$S_N$ deviate from the mean ratio or difference (red horizontal dashed line) by more than the standard error (shaded band). Some large variations in the ratios occurring *e.g.* in ≈ 1954, 1964, 1976, 1985 and 1995 actually appear around the solar minima, when the sunspot numbers are close to zero and relative errors become large. Therefore, those deviations are not significant. On the other hand, Koyama's Wolf numbers are significantly overestimated before 1948. Then, a stable interval follows over the rest of Solar Cycle 18, before significant underestimates occur in the interval 1956-1965 in Solar Cycle 19. The Koyama scale returns to its Solar Cycle 18 level over the whole Solar Cycles 20 and 21, from 1966 to 1983. The scale fluctuates over this interval, but the upward or downward deviations remain less than 6% and do not persist for more than a few years. The only significant excursion is a brief 14% excess in 1977 - 1978, in the rising phase of Solar Cycle 21.

However, by 1983, when Solar Cycle 21 approaches its minimum, the ratio falls to a lower level, with a 9% drop in the counts relative to the pre-1983 numbers. This lower scale continues at a stable level during the next 12 years, until almost the end of Koyama's observations, which finish in the Solar Cycle 21-22 minimum. Note that the final upward deviation of the ratio is not significant, as the low numbers lead to larger relative errors.

Overall, Koyama's $W_K$ series is mainly characterized by temporary fluctuations around a very stable scale, which itself does not change significantly over the interval 1948-1983, *i.e.* over a 35-year time span. Most fluctuations, over timescales longer than 1 year, remain under 6%, with a root-mean-square deviation of 1% on a monthly timescale, and the overall mean $W_K/S_N$ ratio for this long time interval is 0.953 ± 0.005. Only the last 12 years from 1983 to 1995 are marked by a transition to another very stable scale, falling down to 0.876 ± 0.007, thus about 9% below the pre-1983 level. Therefore, if we wish to use the entire series, it is necessary to first raise the values after 1983 by a factor 1.088 ± 0.009, in order to bring them to the same scale as the rest of the series.

The first observations before 1948, were based on a very different setup and must be considered as a separate series (see Table 1). The low number of observations prevents the determination of an accurate correction factor. Therefore, it seems difficult to pre-pend those data to the rest of the series. Contrary to the addition of the last 13 years after 1983, the loss of those few early data is limited, as it would extend the global series by only a few percent in duration.

This global stability is confirmed when deriving the global relation between $W_K$ and $S_N$ by different methods, as shown in Figure 9. Here, we derived the mean ratio, the ratio giving a null mean difference between the two series, a classical linear regression, linear regression forcing the intercept at the origin, and finally a polynomial regression.

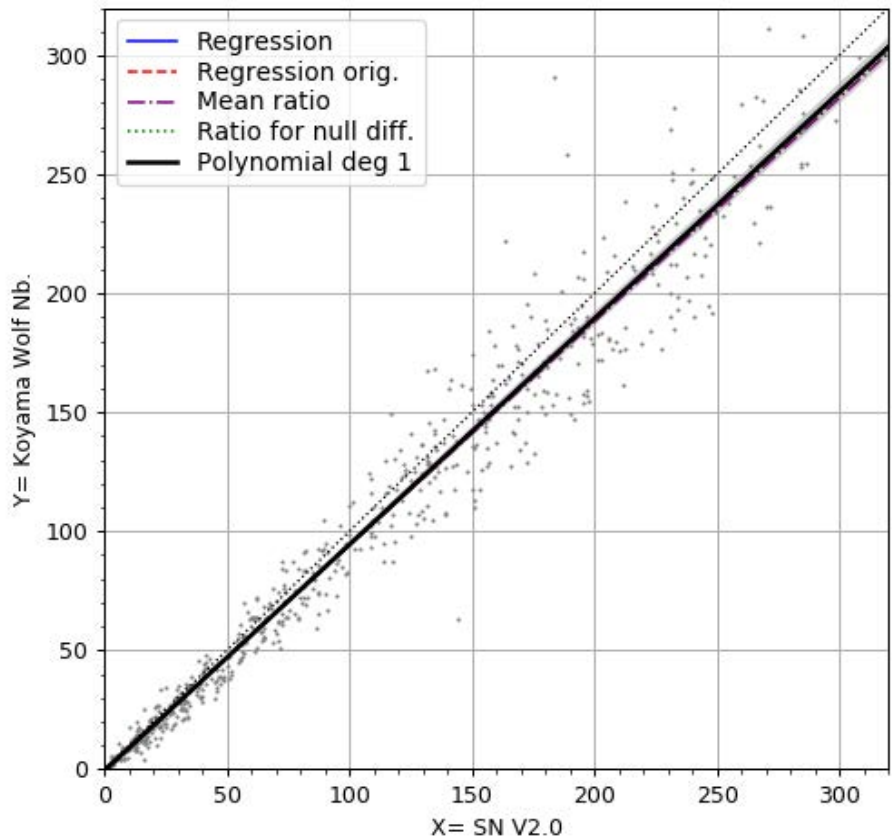

Figure 9: Scatterplot (gray dots) of Koyama's monthly mean Wolf numbers $W_K$ versus the monthly mean sunspot number $S_N$ Version 2.0. Regression lines for five different regression methods are over-plotted (cf. main text). The gray shading corresponds to the interval ± 1 standard deviation from the linear fit with intercept through the origin (hardly visible because it just encompasses all other regression lines). The diagonal dotted line corresponds to a ratio of exactly 1. All regressions match very closely and are below a slope of 1.

All linear regressions give a very high linear correlation coefficient *R* of 0.981. The polynomial fits with degree-two or higher indicate that all terms beyond degree 1 are not significant, and thus that the relation can be considered as fully linear. The resulting linear term matches the other determinations. Therefore, this demonstrates a very high correlation between Koyama's numbers and the reference $S_N$ series, and it also indicates that the relation is fully linear. Taking the linear regression with intercept at the origin as the most representative estimate, the mean $W_K/S_N$ ratio over the entire 50-year series, from 1947 to 1995 equals 0.947 ± 0.005. All other determinations range from 0.941 to 0.951 and thus agree within the 1σ interval.

Taking into account the main transitions identified in the ratios and differences, as mentioned above, we computed the mean $W_K/S_N$ ratios in the sub-intervals between those transitions, in the same way as for the whole series above. The resulting values are listed in Table 2, with each corresponding time window. Over the primary stable interval 1948 to 1983, the $W_K/S_N$ scaling ratio equals 0.953 ± 0.005, thus slightly higher than the overall value given above, by about 0.6%. This is just significant. On the other hand, the final 12 years (1983-1995) give 0.876 ± 0.007, which is 8% below the overall ratio, a largely significant deviation (10 σ). A few other intervals also give significant deviations of the same amplitude but they are much shorter (less than 2 years).

A few $W_K$ monthly means deviate significantly from the corresponding $S_N$ monthly mean. Those few cases correspond to months when Koyama could only observe during very few days. Actually, when she could not use her main telescope(s) for a considerable period, she generally mentioned the reason in her logbook. Table 3 shows the months with a number of observing days below 5 and the identified reason. As there are very few such outliers, they do not change significantly the overall regression. Here, we included all anomalous months, but those few outliers can easily be filtered out if needed.

Table 2: Table of the mean $W_K/S_N$ scaling ratios for the different stable intervals between significant transitions in the mean ratio relative to the sunspot number $S_N$ Version 2.0, taken as reference (here starting from monthly mean k values). The transition dates are best estimates, as they do not occur abruptly from one month to the next in most cases and are superimposed on random fluctuations. The error listed is 1 standard deviation, *R* is the linear correlation coefficient, and the 5[th] column "Rel." lists

the ratio relative to the scale of the primary stable period 1948-1983.

| Time Interval | Ratio | Error | *R* | Rel. | Comment |
|---|---|---|---|---|---|
| 1947-01 to 1995-03 | 0.947 | 0.005 | 0.981 | 0.994 | Full series |
| 1948-01 to 1983-10 | 0.953 | 0.005 | 0.984 | 1 | Primary interval |
| 1947-01 to 1947-12 | 1.118 | 0.053 | 0.683 | 1.173 | Small telescope |
| 1948-01 to 1956-07 | 0.984 | 0.013 | 0.985 | 1.032 | |
| 1956-08 to 1958-03 | 0.897 | 0.014 | 0.984 | 0.941 | |
| 1958-04 to 1959-07 | 0.992 | 0.02 | 0.892 | 1.041 | |
| 1959-08 to 1961-05 | 0.879 | 0.044 | 0.975 | 0.922 | |
| 1961-06 to 1965-12 | 0.898 | 0.015 | 0.992 | 0.942 | |
| 1966-01 to 1977-07 | 0.966 | 0.009 | 0.986 | 1.013 | |
| 1977-08 to 1978-12 | 1.089 | 0.018 | 0.999 | 1.143 | |
| 1979-01 to 1983-10 | 0.948 | 0.009 | 0.983 | 0.995 | |
| 1983-11 to 1995-03 | 0.876 | 0.007 | 0.996 | 0.919 | Final stable period |

Table 3: Period with limited observing days (< 5 days a month) with the corresponding reason, as found in Murayama's notes.

| Period | Reason |
|---|---|
| 1950 Mar | dome repair |
| 1954 Feb – Mar | repair of the equatorial telescope |
| 1960 Mar | dome reparation |
| 1965 Nov -- 1966 Mar | repair of the equatorial telescope |
| 1969 Feb – Mar | hospitalized |
| 1979 Feb | dome repair |

**5.2 Comparison with the original Zürich sunspot number ($S_N$ Version 1.0)**

Before the 2015 re-calibration of the sunspot number series, the original series built primarily by the Zürich Federal Observatory until 1981 had never been modified (Clette and Lefèvre, 2016). This is thus the official reference version that Koyama knew during her observing career.

Our experience with contemporary observers of the SILSO network indicates that in some cases, the observers compare their past counts with the published values of the reference sunspot number, and may be influenced, *e.g.*, by counting tinier marginal sunspots or splitting groups differently in order to raise their personal Wolf numbers if they observe that their numbers tended to be lower than the reference $S_N$ over previous months.

It turns out that several inhomogeneities were diagnosed in the original sunspot number, partly in the 20$^{th}$ century (Clette *et al*., 2014, Clette and Lefèvre, 2016). In particular, a variable drift over up to 20% affected $S_N$ Version 1.0 after 1981, when the Zürich auxiliary station in Locarno, Switzerland, succeeded the Zürich Observatory as pilot station (*e.g.*, Clette *et al*., 2016). Therefore, if by any chance Koyama was influenced by the Zürich-Locarno reference, the flaws affecting the reference may have left an imprint in Koyama's personal scale.

In order to check this, we repeated the calculation of the previous section, but using $S_N$ Version 1.0 as reference in the comparison and the regressions. Figure 10 illustrates the result in exactly the same way as the previous analysis (monthly mean values, Gaussian smoothing for plotted curves, different regressions). Note that before the calculations, $S_N$ Version 1.0 was scaled by a constant factor to match $S_N$ Version 2.0, as the reference unit for the counts changed between the two series (see Clette and Lefèvre, 2016). This is not critical for this diagnostic but it makes visual comparisons easier.

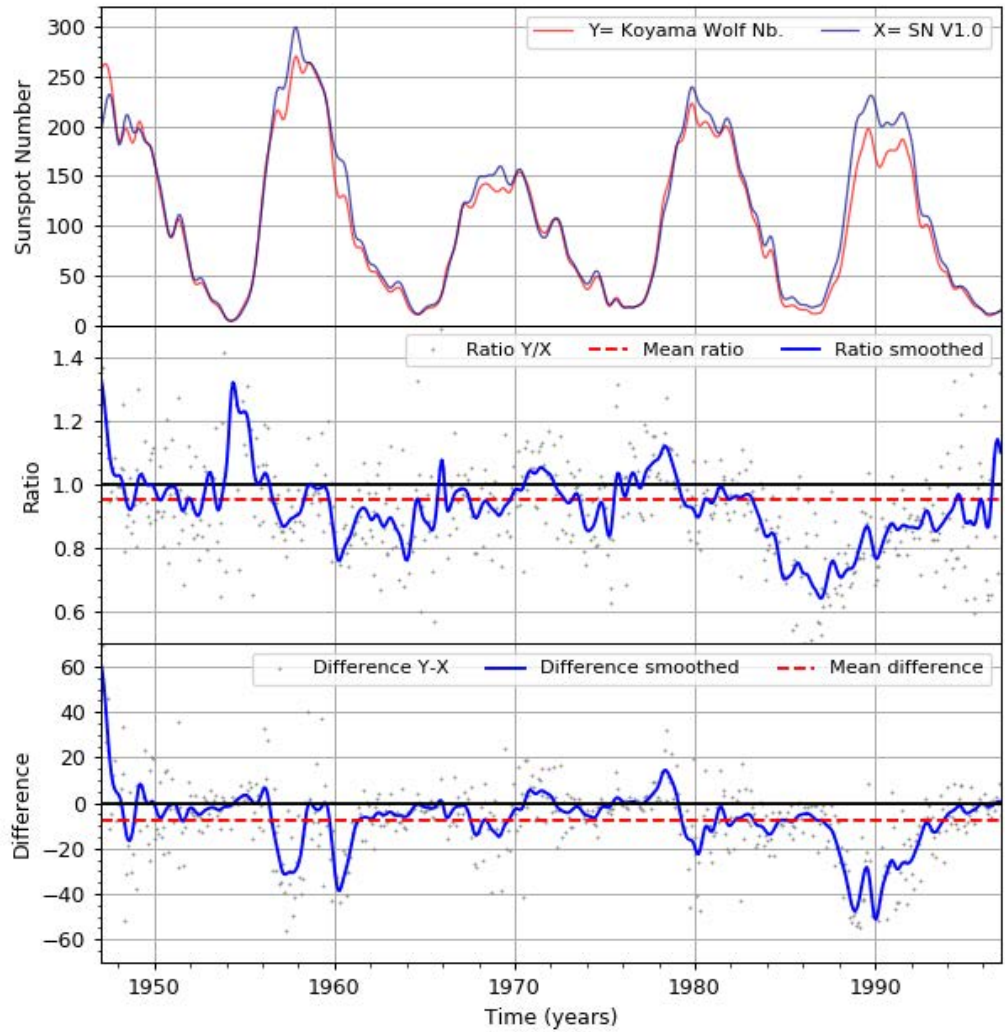

Figure 10: Comparison between the raw Wolf numbers from Koyama $W_K$ and the original sunspot number series $S_N$ Version 1.0, before any of the corrections applied in the 2015 re-calibration. All curves are equivalent to the curves in Figure 8. Here, no error estimates are shown, as the $S_N$ V1.0 did not include error values. A larger and variable deviation occurs after 1983, and is attributable to artificial drifts affecting $S_N$ V1.0 and corrected in V2.0.

The new comparison leads to results that are almost undistinguishable from the results in section 5.1 (Figure 8), though only for years before 1983. The only important difference appears in the period after 1983, thus the interval in which Koyama's numbers switch to a stable but 9% lower scale over the whole interval 1983-1995. Relative to SN Version 1.0, we observe a much deeper deviation by almost 30%, which is variable over the interval. It drifts downwards then upwards, which means that Koyama's numbers dropped even lower relative to the Locarno-based Version 1.0 of the $S_N$ series.

It turns out that this extra deviation matches the diagnosed upward drift of the Locarno pilot station, which resulted in over-counting during those years. Therefore, this test proves that this known flaw in the original SN Version 1.0 series is absent in Koyama's own numbers. We can thus safely conclude that Koyama was not influenced by the Zürich reference. She counted spots and split the groups independently, and the fluctuating inhomogeneities found here were not induced by external factors, beyond the local context of her daily observing routine.

### 5.3 Comparison with the original group sunspot number

As Svalgaard and Schatten (2016) mention a degradation of Koyama's group counts starting in 1981, we repeated the comparison using the original GSN series by Hoyt and Schatten (1998a, 1998b). Indeed, the new backbone series by Svalgaard and Schatten (2016) uses the published part of Koyama data (Koyama, 1985) as a reference for that period, and cannot be used for this comparison, while the original group number is mostly based to Greenwich and the UASF-SOON network data over the Koyama observing window.

We repeated the above analysis, now applying it to Koyama's raw group counts and the Hoyt and Schatten group sunspot number $G_N$. This is illustrated in Figure 11, where the most prominent variations in the ratios and differences found in the SN version 2.0 comparison can be recognized. Smaller random differences in the monthly means can be explained by the following factors:

- Group counts are low numbers compared to the sunspot numbers (factor about 20; see Muñoz-Jaramillo and Vaquero, 2019). The daily numbers are thus distributed over a more limited number of values (quantization effect)
- The base data used to produce daily group numbers come from a smaller number of stations, and the set of stations used for $G_N$ and $S_N$ are non-overlapping.
- The primary reference used for $G_N$ until 1975 is the Greenwich photographic catalog, while the SOON visual counts (network of the US Air Force) were used in replacement after 1975. This may lead to some inhomogeneity in the Hoyt and Schatten series, in the middle of the Koyama time window.

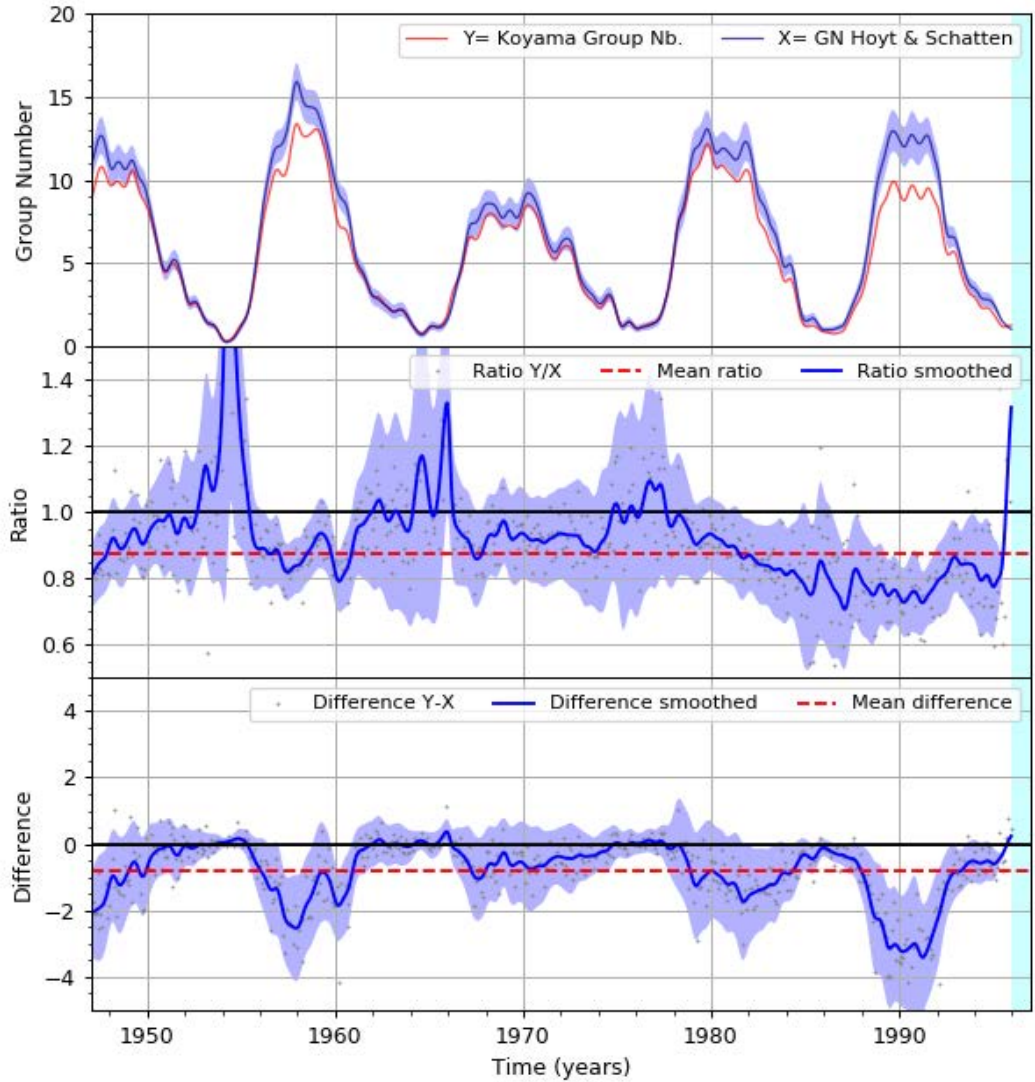

Figure 11: Comparison between the raw group numbers from Koyama $G_K$ and the group sunspot number series $G_N$ produced by Hoyt and Schatten (1998a, 1998b). All curves are equivalent to the curves in Figures 8 and 10. A drift to a lower ratio occurs between 1981 and 1985.

In this alternate comparison, where individual spots are not taken into account, we find again the deviations before 1948, during Solar Cycle 19, and in the last year 1996. We also find a drift to a lower scale during Solar Cycle 21, but here it happens progressively between 1981 and 1985, instead of sharply in 1983. The drift starts earlier in 1981, which matches the time found by Svalgaard and Schatten (2016) when building their new "backbone" group number series.

For group numbers, we find that the mean $G_K/G_N$ ratio during the primary stable period 1948-1981 equals $0.901 \pm 0.004$, and it drops to $0.779 \pm 0.006$ during the period 1985-1995. Koyama's group numbers are thus about 10% lower overall than the reference $G_N$ series, thus a larger difference than for her Wolf numbers. Moreover, the amplitude of the drop is 15%, and is thus significantly larger than for the Wolf number.

This new comparison with an independent series thus confirms the reality of the scale jump around 1981. As we are considering only group counts here, it also indicates that the change of scale is not only due to a change in the spot counts, but is brought in part by the number of groups counted by Koyama. Actually, as the relative amplitude of the transition is larger for groups than for the Wolf numbers, the group counts may actually play a dominant role in the scale change. We also note that Koyama's group counts are smaller by a larger amount (10%) than her Wolf numbers (5%) relative to the reference series over the primary homogeneous part (1948-1983) This again suggests that group counts are probably the main cause of the difference, rather than lower counts of individual spots.

While the spot counts depend almost exclusively on the observer acuity, the group counts can also depend on the way an observer is splitting groups. Therefore, in order to identify the cause of this transition in the Koyama sunspot data, an analysis of her group splitting practices should be carried out. This goes beyond the analysis of the time series of sunspot counts presented here, and requires the examination of her drawings and her synoptic catalogue of individual sunspot groups. Fortunately, the digital version of her catalog is now available for investigating this newly-found issue.

The different timing of the transition (1981 instead of 1983 and sharp versus progressive transition) is probably due to other imperfections in either of the reference series used as comparison. However, as the $S_N$ Version 2.0 series is based on a larger number of independent reference stations (25) than the 1998 GSN and cannot be affected by the Greenwich-SOON transition occurring just a few years before the transition, we consider that the comparison with $S_N$ V2.0 provides the most reliable and accurate diagnostic of Koyama's inhomogeneities.

**6. Interpretation: possible causes of the inhomogeneity**
Based on the fluctuations and transitions revealed by the data series itself, we tried to cross this information with all the changes in Koyama's daily observing routine documented in the logbooks and archives. The most prominent deviations appearing in the data are (*cf.* Table 2): (1) the early steep downward trend in 1947; (2) two "dips" (for ≈ 1 year) around the peak of Solar Cycle 19, and (3) a stable lower level around Solar Cycle 22 (9 % lower than the earlier series).

The initial steep drop in the early phase (≈ 1947) can probably be explained by the initial learning period and the transition from a projection diameter of 10 cm to 30 cm. Moreover, in January 1947, her observations with the new 30 cm projection were carried out for only 7 days, giving a very small statistical sample and large errors on the resulting mean scale.

While the first "dip" around the peak of Solar Cycle 19 falls near the solar maximum in 1958, the other "dip" in 1960 appears during its early declining phase. During this period, Koyama's observations were interrupted due to repairs of the dome slit over almost an entire month (1960 March 4 – 30), and she managed to observe sunspots only for 3 days. This is one of the longest periods without observations for Koyama herself. The small number of observations during this month may have contributed to the "dip" in her observations during this period.

The most significant change in Koyama's observations is probably occurring during Solar Cycle 22, with the stable drop of the scale of her Wolf Numbers by 9%. The $W_K/S_N$ ratio shows a sharp transition around 1983. Nevertheless, we must note that this time falls near a cycle minimum. As a consequence, the lower scale can only be robustly established by 1988, when the activity rises to the next cycle maximum. While Koyama retired from the NSM (current NMNS) in March 1981, she continued to be a museum fellow and to access her main telescope in this museum (Koyama, 1985). It is only after the solar minimum in 1986 that she started to switch more and more often from her main telescope at the Museum to Murayama's 15-cm refractor, with a transition period in 1989 – 1991. As this documented transition involves a change of telescope and observing site, it would be the most likely cause of the shift to lower mean $W_K$ numbers relative to the $S_N$ reference. However, the timing does not match well; as the scale transition comes later than the 1981 retirement and earlier than the 1989 –1991 telescope migration.

Another possible factor is the increasing participation of alternate observers (most probably Sadao Murayama; see Figure 12), who replaced Koyama more often in the late years. It is actually possible to establish when other observers were involved, by consulting the original drawings and logbooks. Indeed, in most sunspot drawings in the Koyama series, the observers' identity is certified by a signature starting in mid-1948, and by personal stamps starting in 1952 (see *e.g.*, lower left of Figure 1). Moreover, some drawings do not include any stamp or signature, most likely indicating that the drawing is not from Koyama, but then not allowing to identify who was the alternate observer. Based on those personal identifiers or lack of identification, we can derive the level of participation of alternate observers over the entire Koyama observing period, except for the early period 1948 – 1951 when stable practices were not yet established for stamps or signatures. This is shown in Figure 13. The most prominent feature in this plot is a significant increase in the participation of replacement observers after 1983. It turns out that this transition in the observing routine coincides with the 1983 jump found in the Wolf numbers, which strongly suggests that both transitions are related. We point out that the influence of those auxiliary observers is not limited to the numerical proportion of drawings made by other observers relative to the genuine drawings made by Koyama herself in the data set. The participation of other observers can also induce a mutual influence on the counts of spots and groups, and may have influenced Koyama's own daily counts even in her personal observations. This indirect influence is very difficult to demonstrate and quantify retrospectively for a team of visual observers. Indeed, by themselves, the documents at our disposal left by Koyama, do not allow us to find a definitive answer. In that sense, the resulting counts and their variations relative to an independent reference, as we calculate above, provide the ground truth for establishing and quantifying the effect. Further diagnostics may help deriving additional evidence, but this would probably involve detailed sunspot catalogs for that epoch, which goes beyond the scope of this study.

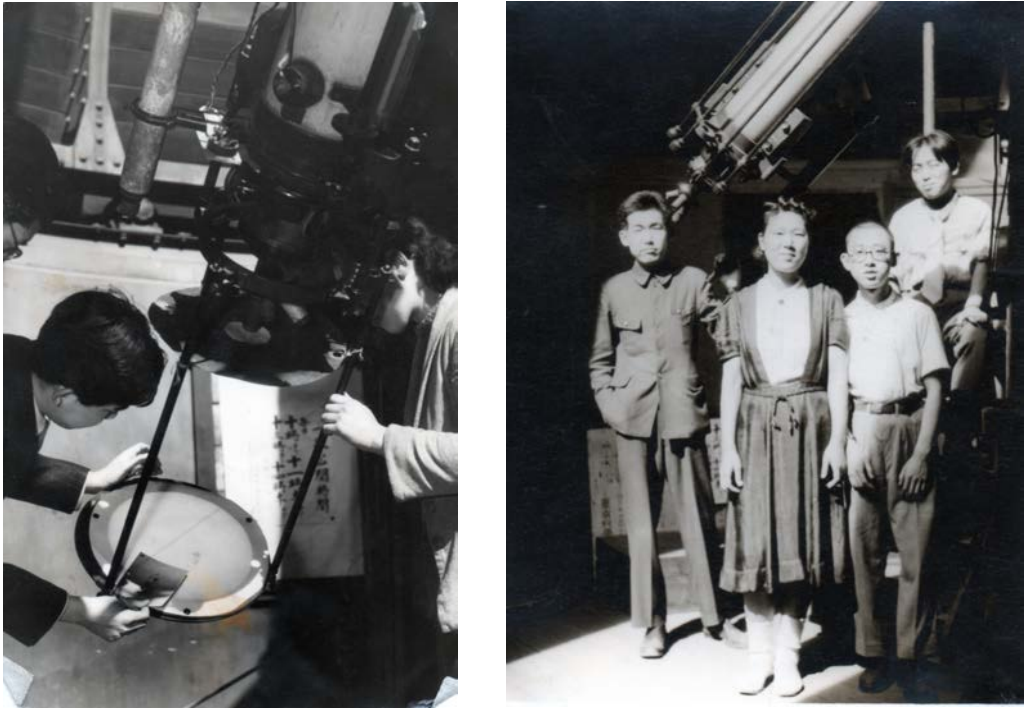

Figure 12: (left) Sadao Murayama and Hisako Koyama using the 20-cm refracting telescope the projecting the solar image with 30-cm projection screen at the NSM.; (right) Hisako Koyama and her colleagues at the observatory dome at Ueno in 1946: Sadao Murayama, Hisako Koyama, Toyokazu Ohtani, and Koichiro Tomita (from left to right). Their signatures are found in the 1946 sunspot drawings (courtesy of Noji Collection).

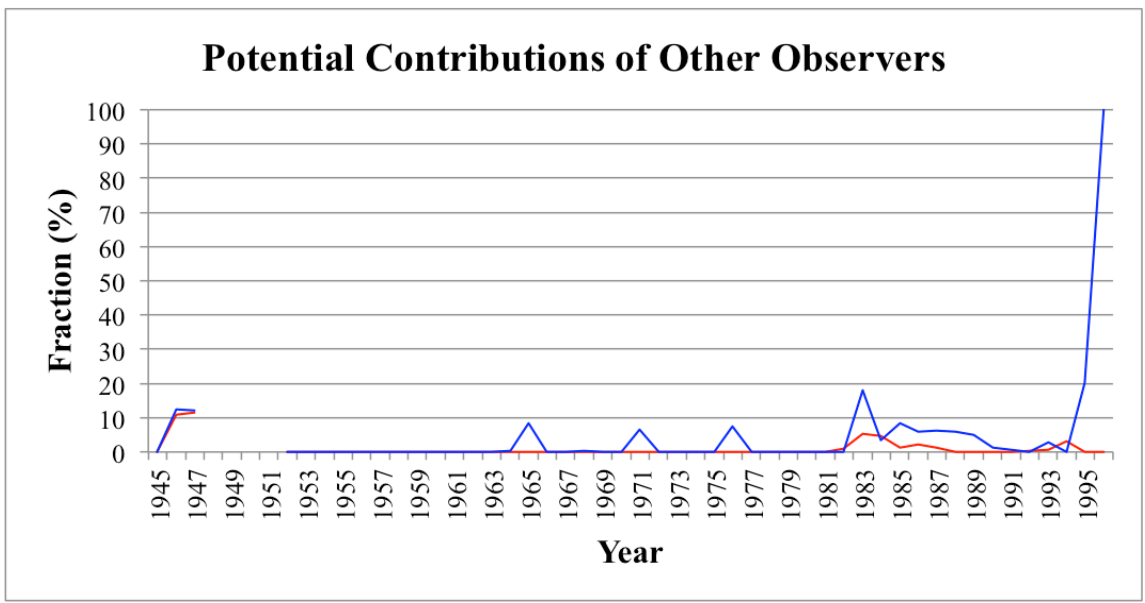

Figure 13: Fraction of potential contributions of other observers – most probably Sadao Murayama's – in Koyama sunspot drawings, on the basis of other observers' signatures (red curve) and lack of stamps/signatures (blue curve). No reliable identifications were available over the interval 1948 – 1951 (see main text).

At this stage, we can at least establish that this transition occurred and affects the homogeneity of the last part of Koyama's data. Fortunately, the diagnosed jump is followed by a second stable period, rather than a continuous and variable drift. Therefore, this last segment of the series remains usable after the application of a simple constant correction factor applied to the whole interval 1983-1995. This correction factor (1.088 ± 0.009; see section 5.1) can be determined with enough accuracy, thanks to the fact that Koyama's numbers remain quite stable during this second interval.

**6. Conclusion:**

In this article, we provided an overview of Hisako Koyama's sunspot observations over her entire observing career, 1945 – 1996, on the basis of her original sunspot drawings and logbooks. While Koyama published her own compilation covering 1947 – 1984 (Koyama, 1985), she started observing already in 1945 and continued until the end of her life in 1996.

Most of Koyama's sunspot drawings and her logbooks have been recently digitized under the database of the NMNS. These logbooks retrace the successive change of main telescope. In the earliest phase[3], she mainly used 3-cm eyepiece observation using the aerial image during 1945 – 1946, and the 20-cm refractor with 10 cm projection during 1946 to 1947 Jan. 21. From 1947 Jan. 22 onward, she used the 20-cm refractor with a larger 30 cm projection, which can be considered as her main standard observing configuration. Finally, she used a 15-cm refractor with 30 cm projection, with a progressive transition over 1989 – 1991.

Consistently observing for more than 150 days per year, she produced unique long-term data series covering more than a half century (1945 – 1996). She recorded not only the total whole-disk sunspot number, but also each hemisphere and the central zone on a daily basis.

---

[3] https://www.kwasan.kyoto-u.ac.jp/~hayakawa/data

The comparison with the SILSO sunspot number series (Version 2.0) does not show long-term drifts within Koyama's data series during the interval 1948-1983. Koyama's Wolf numbers were thus stable, at a level slightly below SN V2.0, which is scaled on the Zürich pilot station. Only temporary deviations occurred, over a timescale of 1 to 3 years, with amplitudes below 10%. However, a jump to a lower scale occurs around 1983, with numbers consistently lower by a constant factor of 0.919 relative to the counts in the preceding main part between 1948 and 1983. In a complementary validation, the comparison with the original SN V1.0 shows that Koyama was not influenced by comparisons between her daily counts and the Zürich reference numbers. Her counts prove to be independent and thus unaffected by known artificial drifts affecting the original SN, as it was published at the epoch of her observations.

The comparison with the original group number confirms a progressive downward transition starting in 1981. This transition thus matches the equivalent transition found in the sunspot numbers, although the exact timing of the transition does not coincide precisely because of other factors present in the group number series. The amplitude of the deviations in the group counts is larger than for the Wolf numbers, indicating that group counts account for a large part to the variation of the Wolf numbers. Understanding how Koyama divided the sunspot groups would thus deserve further attention. This can be studied by returning to her collection of original drawings, which contains the base information. There is thus matter for future deeper studies, which could benefit from the extremely rich information contained in her precious collection of original drawings. Those preserved drawings form the foundation of the entire Koyama heritage.

Looking at shorter timescales, in most cases, we did not find any convincing correspondence between the fluctuations detected in the Koyama sunspot counts and the documented changes in her observing routine. Still, the initial drop in 1947 is most probably due to the initial changes in her observing setup and practices. Likewise, the 1983 jump corresponds with an increased participation of replacement observers over the late periods 1983 – 1996, based on the personal stamps and signatures found on the original drawings. The progressive use of an alternate 15-cm telescope occurred later, by 1989, and was gradual, which did not leave any matching signature in the scale of Koyama's counts.

Overall, Koyama's personal Wolf numbers can be considered as a very stable long-term reference over the 47-year interval 1948 to 1995, provided a single and simple correction is introduced: a constant correction factor 1.088 ± 0.009 applied to all Wolf numbers after 1983. Over shorter intervals, this amazingly long series shows inevitable excursions. However, they remain below a 10% and are mostly of a few percent (1% rms) at a monthly timescale. When used in global statistical reconstructions of sunspot or group number series, these local fluctuations can be filtered out by combining many series of multiple parallel stations, as such fluctuations are essentially random and uncorrelated between the different observers. However, as most of those parallel stations won't offer durations as long as Koyama's series, this is precisely where such an exceptionally long uninterrupted series plays a unique role as a "backbone" connecting epochs separated by half a century, *i.e.*, 4 solar cycles or more, and thus giving the long-term rigidity needed to obtain a fully homogeneous multi-century sunspot-based solar activity index.

**Acknowledgment:**

This work was partly supported by the National Museum of Nature and Science of Japan for its digitization project and the International Space Science Institute (ISSI, Bern, Switzerland) via the International Team 417 "Recalibration of the Sunspot Number Series", chaired by M. Owens and F. Clette. We thank Kazuki Noji for the contemporary photographs of Hisako Koyama, Sadao Murayama, and their colleagues, from the Noji Collection. We thank Leif Svalgaard for his helpful comments. This work and the team of the World Data Center SILSO, which produces and distributes the international sunspot number used in this study, are supported by Belgian Solar-Terrestrial Center of Excellence (STCE) funded by the Belgian Science Policy Office (BelSPo). DJK thanks the National Center for Atmospheric Research (NCAR) for library support. Some of this work is based upon access to the historic High Altitude Observatory Library at the NCAR, which is a major facility sponsored by the National Science Foundation under Cooperative Agreement No. 1852977. This work is also supported by the Grant-in-Aid from the Ministry of Education, Culture, Sports, Science and Technology of Japan, Grant Number JP15H05816 (PI: S. Yoden), and JP17J06954 (PI: H. Hayakawa). HL acknowledges support by JSPS KAKENHI grants 18H01270 (PI: H. Liu), 18H04446 (PI: H. Liu), and 17KK0095 (PI: H. Liu).

Hayakawa et al. (2020) Sunspot Observations by Hisako Koyama: 1945 – 1996, *Monthly Notices of the Royal Astronomical Society*. DOI: 10.1093/mnras/stz3345